# Anisotropic carrier mobility of distorted Dirac cones: theory and application

Ting Cheng,[1,2,#] Haifeng Lang,[1,#] Zhenzhu Li,[2] Zhongfan Liu,[1,2,3] and Zhirong Liu[1,2,3,*]

[1] *College of Chemistry and Molecular Engineering, Peking University, Beijing 100871, China*
[2] *Center for Nanochemistry, Academy for Advanced Interdisciplinary Studies, Peking University, Beijing 100871, China*
[3] *State Key Laboratory for Structural Chemistry of Unstable and Stable Species, Beijing National Laboratory for Molecular Sciences, Peking University, Beijing 100871, China*

[#] Contributed equally to this work.

[*] LiuZhiRong@pku.edu.cn

## Abstract

We have theoretically investigated the intrinsic carrier mobility in semimetals with distorted Dirac cones under both longitudinal and transverse acoustic phonon scattering. An analytic formula for the carrier mobility was obtained. It shows that tilting significantly reduces the mobility. The theory was then applied to 8B-*Pmmn* borophene and borophane (fully hydrogenated borophene), both of which have tilted Dirac cones. The predicted carrier mobilities in 8B-*Pmmn* borophene at room temperature are $14.8 \times 10^5$ and $28.4 \times 10^5$ cm$^2$ V$^{-1}$ s$^{-1}$ along the *x* and *y* directions, respectively, both of which are higher than that in graphene. For borophane, despite its superhigh Fermi velocity, the carrier mobility is lower than that in 8B-*Pmmn* owing to its smaller elastic constant under shear strain.

## I. INTRODUCTION

Since graphene was successfully prepared in 2004 [1], hundreds of two-dimensional (2D)

materials have been reported [2,3], yet single-element 2D materials such as graphene and black phosphorus (BP) are still rare. Recently, three different structures of possible 2D boron monolayer allotropes (called borophene [4,5], which is analogous to graphene) have been successfully grown on Ag(111) surfaces under ultrahigh vacuum conditions [4,5], bringing boron to the forefront of 2D family.

Boron is the nearest neighbor to carbon in the periodic table of elements. Before experimental investigation of borophene, a large number of stable 2D boron monolayer sheets (borophenes) were predicted by first-principles calculations [6-9]. Most of them are metallic [6], including those obtained in experiments [4,5,10]. Others are predicted to be semimetallic with anisotropic Dirac cones, such as *Pmmn* borophene and *P6/mmm* borophene [7,8]. Borophenes are expected to have strong electron-phonon coupling to enable superconducting behavior [11-14]. They possess novel mechanical properties, such as a negative Poisson's ratio and high strength [15-19], and are promising anode materials for high-performance Li ion battery [20-22]. Transition metals can be doped into borophenes to create metallo-borophenes with tunable chemical, magnetic, and optical properties [23-28]. It should be noted that all borophenes achieved in experiments were grown on Ag surfaces [4,5,10,29], and their stability in the free-standing form is questionable. To stabilize the system, surface hydrogenation convert borophene to borophane [30,31], which also possesses a negative Poisson's ratio [32,33]. Interestingly, hydrogenation changes the system from metallic to semimetallic with tilted (distorted) Dirac cones and high Fermi velocity [31]. Among the various pristine borophenes, 8B-*Pmmn* borophene (here "8B" indicates that there are eight boron atoms in the unit cell and "*Pmmn*" specifies the symmetry) is of particularly interest because it is the first borophene predicted to have Dirac cones (which are also tilted) and it is one of the most stable predicted 2D boron allotropes[7]. It has lower formation energy than the borophenes prepared in experiments, which makes it easier to form in the free-standing form. This material has many novel properties. It may open a band gap under an external shear force [34]. Like graphene, the strained lattice in 8B-*Pmmn* induces a giant pseudomagnetic field vector potential and a scalar potential [35].

Dirac materials such as graphene have extraordinary properties and numerous potential applications, many of which are closely related to the ultrahigh carrier mobility of their Dirac cones [36,37]. For example, the room temperature mobility in graphene is as high as $2 \times 10^5$ cm$^2$ V$^{-1}$ s$^{-1}$ after extrinsic disorder is eliminated [38,39], which is two orders of magnitude higher than that of transitional semiconductors such as Si. Electron-phonon scattering is usually considered to be the dominant factor to determine the intrinsic carrier mobility in materials since its influence cannot be eliminated or significantly reduced by improving sample preparation [40,41]. For three-dimensional (3D) and 2D semiconductors, the intrinsic mobility is well described by deformation potential theory [40,42,43], where the band edge energy is shifted by longitudinal acoustic (LA) phonons but not transverse acoustic (TA) phonons. For Dirac materials, a lot of studies have been performed to understand how the mobility is affected by electron-phonon scattering [44-48]. Deformation potential theory has been revised to give an analytic formula for the mobility of Dirac cones. It shows that the effect of TA phonons cannot be ignored in Dirac materials because they cause movement of the Dirac point in reciprocal space [47], and it also explains the previous overestimated deformation potential of graphene in experiments. However, the mobility of tilted Dirac cones has rarely been investigated, and it is unclear whether the tilting of the cones has a significant effect on carrier transportation.

Considering that tilting of Dirac cones has a significant effect on many properties of the material (e.g., it is the major criterion in distinguishing type-I and type-II semimetals [49]), it would be intriguing to study the mobility of tilted Dirac cones.

In this paper, we develop an analytical formula for prediction of the intrinsic carrier mobility of tilted Dirac cones and combine it with first-principles calculations to determine the properties of semimetallic borophene and borophane.

## II. THEORETICAL ANALYSIS OF THE MOBILITY OF TILTED DIRAC CONES

In this section, we are concerned with the effect of tilting of Dirac cones on the mobility, and we develop an analytical formula for the mobility under acoustic phonon scattering conditions. It is an extension of deformation potential theory for untilted Dirac cones [47,50].

### A. Hamiltonian of tilted Dirac cones and the response to strain

Without loss of generality, we assume that a Dirac cone is tilted in the $y$ direction. The effective Hamiltonian can then be expanded around the Dirac point to a linear term of the wave vector **k** (measured with respect to the unstrained Dirac point):

$$\mathbf{H_k} = \begin{bmatrix} e_{on} + \hbar\upsilon_t k_y & \hbar\upsilon_F(k_x - ik_y) \\ c.c. & e_{on} + \hbar\upsilon_t k_y \end{bmatrix}, \quad (1)$$

where $e_{on}$ is the on-site energy, $\upsilon_F$ is the isotropic Fermi velocity when the cone is not tilted, and $\upsilon_t$ is the change of velocity owing to the tilting effect. The eigenenergy and the eigenfunction are solved to be

$$E_\mathbf{k} = e_{on} + \hbar\upsilon_t k_y \pm \hbar\upsilon_F|\mathbf{k}|, \quad (2)$$

and

$$\psi_{\mathbf{k}} = \frac{1}{\sqrt{2}} \begin{pmatrix} \pm e^{-i\theta_{\mathbf{k}}} \\ 1 \end{pmatrix}, \tag{3}$$

respectively, with the corresponding carrier velocity

$$\mathbf{v}_{\mathbf{k}} = \left( \upsilon_F \cos\theta_{\mathbf{k}}, \upsilon_F \sin\theta_{\mathbf{k}} \pm \upsilon_t \right), \tag{4}$$

where $\theta_{\mathbf{k}}$ is the angle of $\mathbf{k}$ in the $xy$ plane. Equation (4) suggests that carriers running in the $y$ direction have two different velocities, $\upsilon_F+\upsilon_t$ and $\upsilon_F-\upsilon_t$, owing to tilting of the cone.

The effect of acoustic phonons can be described in terms of small uniaxial/shear strain. For the materials with Dirac cones, the Dirac cones will move to a new location and preserve a cone-like structure under a small strain [47]. The shifting of the Dirac point will leave a pseudo gap opening at the original Dirac point location and the energy level at the Dirac point will be lifted. Generally, the linear response of the Hamiltonian with respect to a strain $\varepsilon_{\text{strain}}$ can be expressed as

$$\Delta \mathbf{H}_{\mathbf{k}} = \begin{bmatrix} S_1 & S_{12} \\ \text{c.c.} & S_1 \end{bmatrix} \varepsilon_{\text{strain}} = \begin{bmatrix} S_1 & -E_\beta e^{-i\theta_{\text{move}}} \\ \text{c.c.} & S_1 \end{bmatrix} \varepsilon_{\text{strain}}, \tag{5}$$

where $S_1$ is a real number, and $S_{12}$ is a complex number which is further expressed into $-E_\beta e^{-i\theta_{\text{move}}}$ ($E_\beta \geq 0$ without loss of generality) for convenience. The eigenenergy under strain is then

$$E_{\mathbf{k}} = e_{\text{on}} + \hbar \upsilon_t k_y + S_1 \varepsilon_{\text{strain}} \pm \hbar \upsilon_F \left| \mathbf{k} - \mathbf{k}_{\text{move}} \right|, \tag{6}$$

where $\mathbf{k}_{\text{move}}$ is the new location of the Dirac point given by

$$\mathbf{k}_{\text{move}} = \frac{E_\beta}{\hbar \upsilon_F} e^{i\theta_{\text{move}}} \varepsilon_{\text{strain}}. \tag{7}$$

In other words, the Dirac point is shifted by a distance of $E_\beta \varepsilon_{\text{strain}} / \hbar \upsilon_F$ along the direction with angle $\theta_{\text{move}}$. Therefore, the energy level at the new Dirac point (DP) is

$$E_{\text{DP}} = E_{\mathbf{k}}(\mathbf{k} = \mathbf{k}_{\text{move}}) = e_{\text{on}} + \left(S_1 + \frac{\upsilon_{\text{t}}}{\upsilon_{\text{F}}} E_\beta \sin\theta_{\text{move}}\right)\varepsilon_{\text{strain}}. \tag{8}$$

A pseudogap opens at the original Dirac point location:

$$E_{\text{gap}}(\mathbf{k} = 0) = 2E_\beta \varepsilon_{\text{strain}}, \tag{9}$$

giving a convenient relation for practical calculations

$$E_\beta = \frac{1}{2}\frac{\partial E_{\text{gap}}(\mathbf{k}=0)}{\partial \varepsilon_{\text{strain}}}. \tag{10}$$

$E_\beta$ is usually called the deformation hopping constant because it is related to the response of the hopping term between the two bases with respect to the strain [47]. Similar to treatment of conventional deformation potential theory, we introduce the deformation potential constant ($E_1$) to describe the variation of the band-edge energy:

$$E_1 \equiv \frac{\partial E_{\text{DP}}}{\partial \varepsilon_{\text{strain}}} = S_1 + \frac{\upsilon_{\text{t}}}{\upsilon_{\text{F}}} E_\beta \sin\theta_{\text{move}}. \tag{11}$$

In tilted Dirac cones, $E_1$ is not only related to the diagonal element of the perturbation Hamiltonian [Eq. (5)], but is also dependent on the off-diagonal element. With Eqs. (6), (7), (10), and (11), parameters such as $E_1$, $E_\beta$, $S_1$ and $\theta_{\text{move}}$ can be readily extracted from first-principles calculations.

## B. Mobility of tilted Dirac cones

We now consider the carrier scattering caused by acoustic phonons, whose effects are described in terms of small strains. With Eqs. (3) and (5), the transition matrix element between states **k** and **k′** for electrons under perturbation is determined to be

$$M_{\mathbf{k}',\mathbf{k}} = \langle \psi_{\mathbf{k}'} | \Delta \mathbf{H} | \psi_{\mathbf{k}} \rangle = \varepsilon_{\text{strain}} \left[ S_1 \cos\frac{\theta_{\mathbf{k}'} - \theta_{\mathbf{k}}}{2} - E_\beta \cos\left(\frac{\theta_{\mathbf{k}'} + \theta_{\mathbf{k}}}{2} - \theta_{\text{move}}\right) \right] e^{i\frac{\theta_{\mathbf{k}'} - \theta_{\mathbf{k}}}{2}}, \tag{12}$$

with the phonon wave vector $\mathbf{q} = \mathbf{k}' - \mathbf{k}$. For 2D materials with Dirac cones, the moving direction of a Dirac point rotates approximately twice as fast as the rotating strain, which is further related to the direction of **q** [47,51-53]:

$$\theta_{move} \approx -2\theta_{\mathbf{q}} + \text{constant}. \tag{13}$$

The strain is related to the amplitude ($u_\mathbf{q}$) of phonon vibration as $\varepsilon_{strain} = u_\mathbf{q} q$, so

$$M_{\mathbf{k'},\mathbf{k}} \approx u_\mathbf{q} q \left[ S_1 \cos\frac{\theta_{\mathbf{k'}} - \theta_\mathbf{k}}{2} - E_\beta \cos\left(\frac{3(\theta_{\mathbf{k'}} + \theta_\mathbf{k})}{2} - \theta_0\right) \right] e^{i\frac{\theta_{\mathbf{k'}} - \theta_\mathbf{k}}{2}}, \tag{14}$$

where $\theta_0$ is an unimportant constant.

When the temperature is much higher than the characteristic degenerate temperature [47], with Fermi's golden rule and the equipartition principle, the probability of scattering between **k** and **k′** under LA phonons becomes

$$W_{\mathbf{k'},\mathbf{k}}^{(LA)} = \frac{2\pi k_B T}{\hbar C_{11} A} \left[ S_1 \cos\frac{\theta_{\mathbf{k'}} - \theta_\mathbf{k}}{2} - E_\beta \cos\left(\frac{3(\theta_{\mathbf{k'}} + \theta_\mathbf{k})}{2} - \theta_0\right) \right]^2 \delta(\varepsilon_{\mathbf{k'}} - \varepsilon_\mathbf{k}), \tag{15}$$

where $\varepsilon_\mathbf{k}$, $\varepsilon_{\mathbf{k'}}$ are the energies of the **k** and **k′** states, respectively, $C_{11}$ is the elastic modulus, and $A$ is the area of the 2D sample. Both emission and absorption of phonons are considered. Here, isotropic $S_1$ and $E_\beta$ are assumed, while the anisotropic case will be discussed below. The relaxation time is defined as

$$\frac{1}{\tau_\mathbf{k}} = \frac{A}{(2\pi)^2} \int W_{\mathbf{k'},\mathbf{k}}^{(LA)} \left(1 - \frac{\mathbf{v}_\mathbf{k} \cdot \mathbf{v}_{\mathbf{k'}}}{|\mathbf{v}_\mathbf{k}|^2}\right) d^2\mathbf{k'}, \tag{16}$$

which can be calculated based on Eqs. (2), (4), and (15). Incorporation of TA phonons is straightforward, as will be shown. The rigorously analytical solution is too complicated (see Supporting Information), so we give the result for second-order expansion of the tilting factor ($\alpha_t \equiv v_t / v_F$):

$$\frac{1}{\tau_\mathbf{k}} = \frac{k_B T k}{4\hbar^2 C_{11} v_F} \left\{ (S_1^2 + 2E_\beta^2) + 2\alpha_t E_\beta^2 \sin\theta_\mathbf{k} + 2\alpha_t^2 \left[\frac{1}{2}(2S_1^2 + 3E_\beta^2) + \frac{5}{8} S_1^2 \cos(2\theta_\mathbf{k}) + \frac{3}{8} E_\beta^2 \cos(4\theta_\mathbf{k} - \theta_0)\right] \right\}$$

,(17)

which is also helpful in the forthcoming determination of the mobility. The residual error owing to expansion is small (see Supporting Information). The zero-order term is consistent with the result of the untilted case [47]. The solution for the mobility is based on the Boltzmann equation [54], where the mobility can be determined by

$$\mu = \frac{\sigma}{ne} = \frac{e\int \tau_{\mathbf{k}} \frac{\partial n_F}{\partial \varepsilon} \upsilon_{\mathbf{k}}\upsilon_{\mathbf{k}} \mathrm{d}^2\mathbf{k}}{\int n_F \mathrm{d}^2\mathbf{k}}, \quad (18)$$

where $\sigma$ is the conductivity, $n$ is the carrier density, and $n_F$ is the equilibrium Fermi-Dirac distribution. Inserting Eqs. (4) and (17) into Eq. (18), at the neutrality point (NP) where the Fermi level $E_F = e_{\mathrm{on}}$, the exact result is also expanded to the second-order of $\alpha_t$ owing to the complexity of the result. Finally, we obtain the carrier mobility of tilted Dirac cones:

$$\begin{cases} \mu_{x,\mathrm{NP}} = \mu_{x,\mathrm{NP}}^{(0)}\left(1-\alpha_t^2\right)^{\frac{3}{2}}\left[1 - \frac{19S_1^4 + 54S_1^2 E_\beta^2 + 24E_\beta^4}{8\left(S_1^2 + 2E_\beta^2\right)^2}\alpha_t^2\right] \\ \mu_{y,\mathrm{NP}} = \mu_{y,\mathrm{NP}}^{(0)}\left(1-\alpha_t^2\right)^{\frac{3}{2}}\left[1 - \frac{5S_1^4 + 42S_1^2 E_\beta^2 + 40E_\beta^4}{8\left(S_1^2 + 2E_\beta^2\right)^2}\alpha_t^2\right] \end{cases}, \quad (19)$$

where the factor $\left(1-\alpha_t^2\right)^{\frac{3}{2}}$ comes from the effect of tilting on the carrier density $n$, and $\mu_{\mathrm{NP}}^{(0)}$ is the corresponding result for untilted Dirac cones:

$$\mu_{x,\mathrm{NP}}^{(0)} = \mu_{y,\mathrm{NP}}^{(0)} = \frac{e\hbar^3 \upsilon_F^4 C_{11}}{0.82(k_B T)^3}\left[S_1^2 + 2E_\beta^2\right]^{-1}. \quad (20)$$

The odd-order terms of $\alpha_t$ disappear owing to the symmetry. The residual error owing to expansion is negligible, as will be discussed in Fig. 5(b). Tilting of the Dirac cone in the *y* direction affects the mobility in both the *x* and *y* directions, and it always decreases the mobility compared with the corresponding untilted case. When the system is doped (i.e., $|E_F - e_{\mathrm{on}}| \ll k_B T$), the carrier density *n* is a basic parameter and the mobility is

$$\begin{cases} \mu_{x,\text{doped}} = \mu_{x,\text{doped}}^{(0)} \left[ 1 - \frac{19S_1^4 + 54S_1^2 E_\beta^2 + 24E_\beta^4}{8\left(S_1^2 + 2E_\beta^2\right)^2} \alpha_t^2 \right] \\ \mu_{y,\text{doped}} = \mu_{y,\text{doped}}^{(0)} \left[ 1 - \frac{5S_1^4 + 42S_1^2 E_\beta^2 + 40E_\beta^4}{8\left(S_1^2 + 2E_\beta^2\right)^2} \alpha_t^2 \right] \end{cases}, \quad (21)$$

with the untilted value

$$\mu_{x,\text{doped}}^{(0)} = \mu_{y,\text{doped}}^{(0)} = \frac{4e\hbar v_F^2 C_{11}}{\pi k_B T n} \left[ S_1^2 + 2E_\beta^2 \right]^{-1}. \quad (22)$$

Until now, we only considered the contribution of LA phonons. When both LA and TA phonons are present, their contributions to $1/\tau_\mathbf{k}$ in Eqs. (16) and (17) are additive. TA phonons (shear strain) have no contribution to $S_1$ owing to the symmetry. Therefore, to incorporate the effect of TA phonons, we can simply make the following replacement in Eqs. (19-22):

$$E_\beta^{\,2} \to E_{\beta,11}^{\,2} + E_{\beta,66}^{\,2} \frac{C_{11}}{C_{66}}, \quad (23)$$

which is similar to the case of graphene [47]. $E_{\beta,11}$ and $E_{\beta,66}$ are the deformation hopping constants under uniaxial and shear strains, respectively, and $C_{11}$ and $C_{66}$ are the corresponding elastic moduli. (It should be noted that according to the Voigt notation, $C_{66}$ and $E_{\beta,66}$ should be used instead of $C_{44}$ and $E_{\beta,44}$ used in a previous study [47].)

### C. Tilted elliptic cone

When an elliptic cone (with anisotropic Fermi velocity $v_{Fx} \neq v_{Fy}$) instead of a regular cone ($v_{Fx} = v_{Fy}$) is tilted, the effective Hamiltonian can be written as

$$\mathbf{H}_\mathbf{k} = \begin{bmatrix} e_\text{on} + \hbar v_t k_y & \hbar v_{Fx} k_x - i\hbar v_{Fy} k_y \\ \text{c.c.} & e_\text{on} + \hbar v_t k_y \end{bmatrix}. \quad (24)$$

The eigenenergy is

$$E_\mathbf{k} = e_\text{on} + \hbar v_t k_y \pm \hbar \sqrt{\left(v_{Fx} k_x\right)^2 + \left(v_{Fy} k_y\right)^2}. \quad (25)$$

Under the linear response of the Hamiltonian [Eq. (5)], the deformation potential constant ($E_1$) to

describe the variation of the band-edge (Dirac point) energy is given by

$$E_1 \equiv \frac{\partial E_{DP}}{\partial \varepsilon_{strain}} = S_1 + \frac{v_t}{v_{Fy}} E_\beta \sin\theta_{move}, \tag{26}$$

where $v_{Fy}$ is used to replace $v_F$ in Eq. (11). The relation between the deformation hopping constant $E_\beta$ and the opened gap at the original Dirac point location

$$E_\beta = \frac{1}{2} \frac{\partial E_{gap}(\mathbf{k}=0)}{\partial \varepsilon_{strain}}, \tag{27}$$

remains identical to Eq. (10).

Using the following transformation:

$$\begin{cases} \tilde{k}_x = v_{Fx} k_x \\ \tilde{k}_y = v_{Fy} k_y \end{cases}, \tag{28}$$

the Hamiltonian in Eq. (24) "returns" to the isotropic form

$$\mathbf{H}_\mathbf{k} = \begin{bmatrix} e_{on} + \hbar \alpha_t \tilde{k}_y & \hbar(\tilde{k}_x - i\tilde{k}_y) \\ c.c. & e_{on} + \hbar \alpha_t \tilde{k}_y \end{bmatrix}, \tag{29}$$

where the tilting factor is redefined as

$$\alpha_t \equiv \frac{v_t}{v_{Fy}}. \tag{30}$$

Analysis of the mobility can be performed in a similar way as the previous Section (see Supporting Information for details). As an approximate result, the mobility in Eqs. (19) and (21) remains valid with redefined $\alpha_t$ in Eq. (25) and the revised untilted values

$$\begin{cases} \mu_{x,NP}^{(0)} = \frac{e\hbar^3 v_{Fx}^3 v_{Fy} C_{11}}{0.82(k_B T)^3} \left[S_1^2 + 2E_\beta^2\right]^{-1} \\ \mu_{y,NP}^{(0)} = \frac{e\hbar^3 v_{Fx} v_{Fy}^3 C_{11}}{0.82(k_B T)^3} \left[S_1^2 + 2E_\beta^2\right]^{-1} \end{cases} \tag{31}$$

and

$$\begin{cases} \mu_{x,\text{doped}}^{(0)} = \dfrac{4e\hbar v_{Fx}^2 C_{11}}{\pi k_B T n}\left[S_1^2 + 2E_\beta^2\right]^{-1} \\ \mu_{y,\text{doped}}^{(0)} = \dfrac{4e\hbar v_{Fy}^2 C_{11}}{\pi k_B T n}\left[S_1^2 + 2E_\beta^2\right]^{-1} \end{cases} \quad (32)$$

for elliptic cone. More rigorous and lengthy results were also obtained (see Supporting Information), but the improvement in the accuracy is limited, supporting the validity of Eqs. (19), (21), and (30)–(32) for tilted elliptic cones.

When both $S_1$ and $E_\beta$ are anisotropic, it has been proved that the following replacement

$$\begin{cases} S_1^2 \to \left(\dfrac{S_{1x}+S_{1y}}{2}\right)^2 + \dfrac{1}{2}\left(\dfrac{S_{1x}-S_{1y}}{2}\right)^2 \\ E_\beta^2 \to \left(\dfrac{E_{\beta x}+E_{\beta y}}{2}\right)^2 + \dfrac{1}{2}\left(\dfrac{E_{\beta x}-E_{\beta y}}{2}\right)^2 + \dfrac{C_{11}}{C_{66}}E_{\beta,66}^2 \end{cases} \quad (33)$$

is applicable for untilted cones[47], where the subscripts "$x$" and "$y$" indicate that the values are measured under the uniaxial strain along the $x$ and $y$ directions. If tilting occurs in an elliptic cone with anisotropic $S_1$ and $E_\beta$, all of these anisotropic factors entangle together and it is too complex to achieve an analytical solution. Therefore, we directly used Eq. (33) as an approximation.

### III. NUMERICAL RESULTS FOR BOROPHENE AND BOROPHANE

In this section, combined with first-principles calculations, the theoretical formalism developed above is applied to borophene and borophane with tilted Dirac cones to investigate their electronic response to strain and the resulting mobility.

#### A. Calculation method

We implemented density functional theory (DFT) calculations with the Vienna ab initio simulation package (VASP) [55,56]. Geometric optimization and electronic structure calculations under different uniaxial and shear strain were performed. We adopted plane-wave basis sets, the projector augmented wave (PAW) pseudopotential, and the Perdew-Burke-Ernzerhof (PBE) exchange correlation function [57,58]. The structure was optimized with a 45 × 45 × 1 Monkhorst–Pack *k*–mesh [59] and a cutoff kinetic energy of 450 eV. The structure was fully relaxed until all of the atomic forces were less than 0.01 eV/Å. Interactions between adjacent layers were limited by setting vacuum intervals of at least 13 Å and 21Å for borophene and borophane, respectively.

### B. Atomic and band structures

The systems we investigated are 8B-*Pmmn* borophene and 2BH-*Pmmn* borophane (Fig. 1). Similar to the notation 8B-*Pmmn* for borophene, "2BH" is used to indicate that there are two boron atoms and two hydrogen atoms in the unit cell and "*Pmmn*" is the symmetry of borophane. Both 8B-*Pmmn* and 2BH-*Pmmn* have a rectangle unit cell, rather than the hexagonal unit cell of graphene. 8B-*Pmmn* has two nonequivalent B atoms in its unit cell according to the symmetry, which are shown by different colors in Fig. 1(a). For 2BH-*Pmmn*, all of the B atoms are equivalent related with symmetry, as are all of the H atoms. The optimized lattice constants were determined to be $a$ = 4.52 Å and $b$ = 3.26 Å for 8B-*Pmmn*, and $a$ = 2.83 Å and $b$ = 1.94 Å for 2BH-*Pmmn*. All of the results are in line with previous studies[7,31-33].

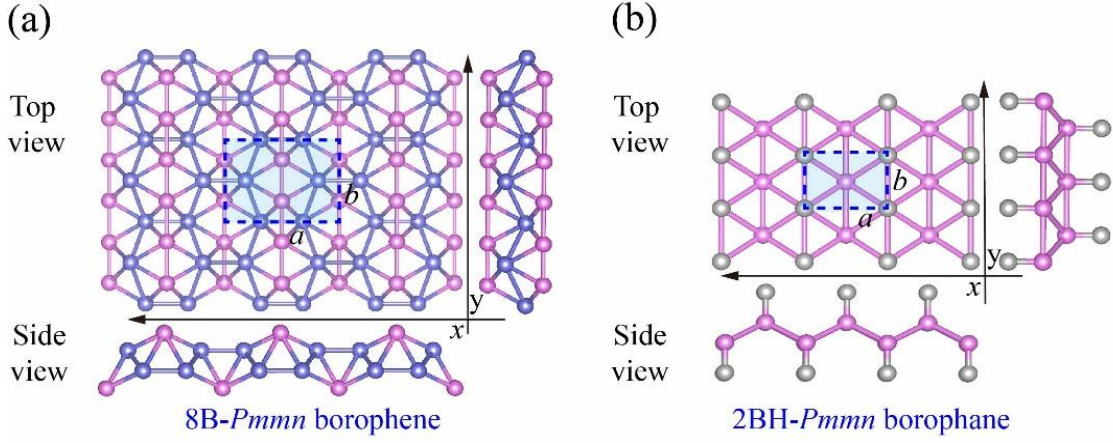

FIG. 1. Top and side views of (a) 8B-*Pmmn* borophene and (b) 2BH-*Pmmn* borophane. In (a), the two different colors distinguish the two types of nonequivalent B atoms. In (b), the pink and gray balls represent B and H atoms, respectively. The unit cells are indicated by dashed shaded rectangles.

The calculated band structures of 8B-*Pmmn* and 2BH-*Pmmn* are shown in Fig. 2. For both systems, the valence and conduction bands near the Fermi level meet at a single point (Dirac point) along the *k*-lines from Γ to X. The group velocity is anisotropic. The band structure near the Dirac point is well described by Eq. (25), and the fitting for 8B-*Pmmn* gives the following parameters: $v_{Fx} = 5.34 \times 10^5$ m/s, $v_{Fy} = 7.85 \times 10^5$ m/s, and $v_t = -3.45 \times 10^5$ m/s. Therefore, the tilting factor of the Dirac cone in 8B-*Pmmn* is $\alpha_t \equiv v_t / v_{Fy} = -0.44$. The resulting tilted Fermi velocities in the *y* direction are thus $v_{Fy} + v_t = 4.4 \times 10^5$ m/s and $v_{Fy} - v_t = 11.3 \times 10^5$ m/s, agreeing well with the previous calculation on 8B-*Pmmn* (4.6 and $11.6 \times 10^5$ m/s) [7].

For 2BH-*Pmmn* borophane, the fitting of the tilted Dirac cone with Eq. (25) gives $v_{Fx} = 7.7 \times 10^5$ m/s, $v_{Fy} = 13.48 \times 10^5$ m/s, and $v_t = -3.86 \times 10^5$ m/s. The tilting factor is $\alpha_t = -0.29$, which is smaller than that of 8B-*Pmmn*. The resulting tilted Fermi velocities in the *y* direction are thus $v_{Fy} + v_t = 9.62 \times 10^5$ m/s and $v_{Fy} - v_t = 17.34 \times 10^5$ m/s, or, 40 eV Å and 72 eV Å, respectively, converted into the slopes of the two linear dispersion bands in the *y* direction using $\partial E_k/\partial k = h(v_{Fy} \pm v_t)$ (noted that the 2π factor is usually ignored in determing *k* from the unit constants) [7,60]). The values in unit of eV Å are practically identical to previously reported values[31]. Reference [32] overestimated the tilted Fermi velocity in unit of m/s in the conversion. The tilted Fermi velocity here is about twice as large as that in graphene (8.2 × 10$^5$ m/s, 34 eV Å) [7,60]. The Fermi level of 2BH-*Pmmn* is −4.61 eV, which is higher than that of 8B-*Pmmn* (−5.12 eV) because of the stabilization effect of surface hydrogenation. For comparison, the Fermi level of graphene is −4.26 eV.

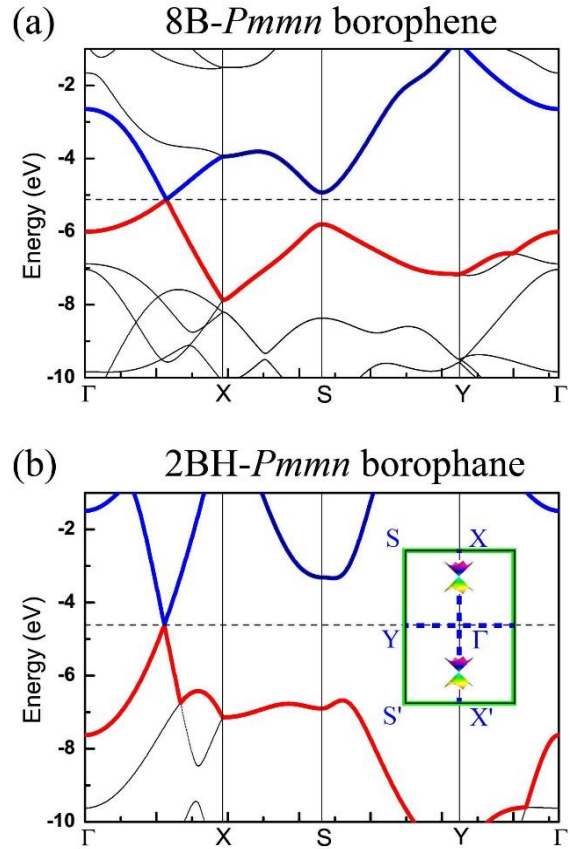

FIG. 2. Band structure of (a) 8B-*Pmmn* borophene and (b) 2BH-*Pmmn* borophane. The Dirac cones are located where the conduction band (blue line) and the valence band (red line) meet, which is (0, 0.2953) for 8B-*Pmmn* and (0, 0.2879) for 2BH-*Pmmn* (measured in the fractional coordinates). The energies were measured with respect to the vacuum energy. The Fermi levels

are plotted in dashed line. The first Brillouin zone with colored cones is shown in the insert of (b).

### C. Stain effect on Dirac cones

The applied homogeneous strain can generally be written as a second-order tensor:

$$\boldsymbol{\varepsilon} = \begin{bmatrix} \varepsilon_{xx} & \gamma \\ 0 & \varepsilon_{yy} \end{bmatrix}, \qquad (29)$$

where $\varepsilon_{xx}$ and $\varepsilon_{yy}$ are the uniaxial strain along the $x$ and $y$ directions, respectively, and $\gamma$ is the engineering shear strain. The lattice vector after strain is $\mathbf{r} = (\mathbf{I} + \boldsymbol{\varepsilon})\mathbf{r_0}$, where $\mathbf{I}$ is the unit matrix and $\mathbf{r_0}$ is the undeformed lattice vector. Three types of strain in the range of ±4% were separately applied.

The band structure of 8B-$Pmmn$ near the Dirac point under uniaxial strain $\varepsilon_{yy}$ = 2% is shown in Fig. 3 to compare with that of the unstrained system. The cone-like structure is well preserved under strain (i.e., no gap opens). However, the Dirac point shifts from (0, 0.2953) to a new location (0, 0.2867), leaving a pseudogap of about 0.16 eV at the original Dirac point location. The Fermi level (the band energy at the Dirac point, $E_{DP}$) also increases from −5.12 eV to −5.07 eV. Uniaxial strain does not change the $Pmmn$ symmetry, and the Dirac point is located at the Γ–X line protected by the mirror symmetry. Therefore, uniaxial strain makes the Dirac point move along the Γ–X line, but it cannot make it deviate from the line. For shear strain $\gamma$, denoting the displacement of the Dirac point as ($\Delta k_x$, $\Delta k_y$), the mirror reflection of $x \rightarrow -x$ would transform $\gamma$ to $-\gamma$ and the displacement of the Dirac point to ($-\Delta k_x$, $\Delta k_y$). Because action of shear strain $\gamma$ together with shear strain $-\gamma$ should restore the system to the unstrained state, we have $\Delta k_y$ = 0 for shear strain (i.e., the movement direction of the Dirac point under shear strain is perpendicular to the Γ–X line), being consistent with the numerical result (see Supporting Information).

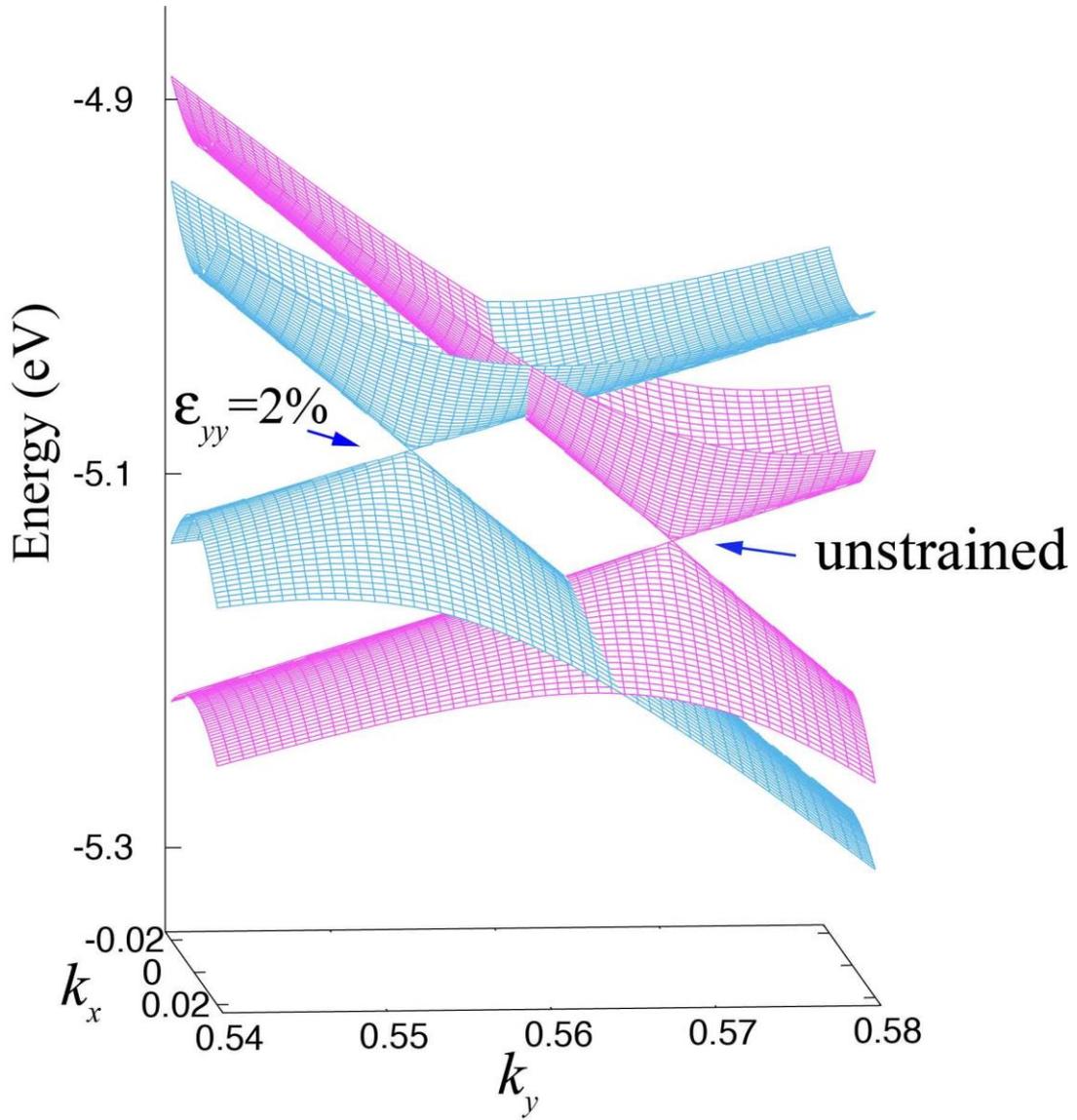

FIG. 3. Strain effect on the band structure of 8B-*Pmmn* borophene near the Dirac point.

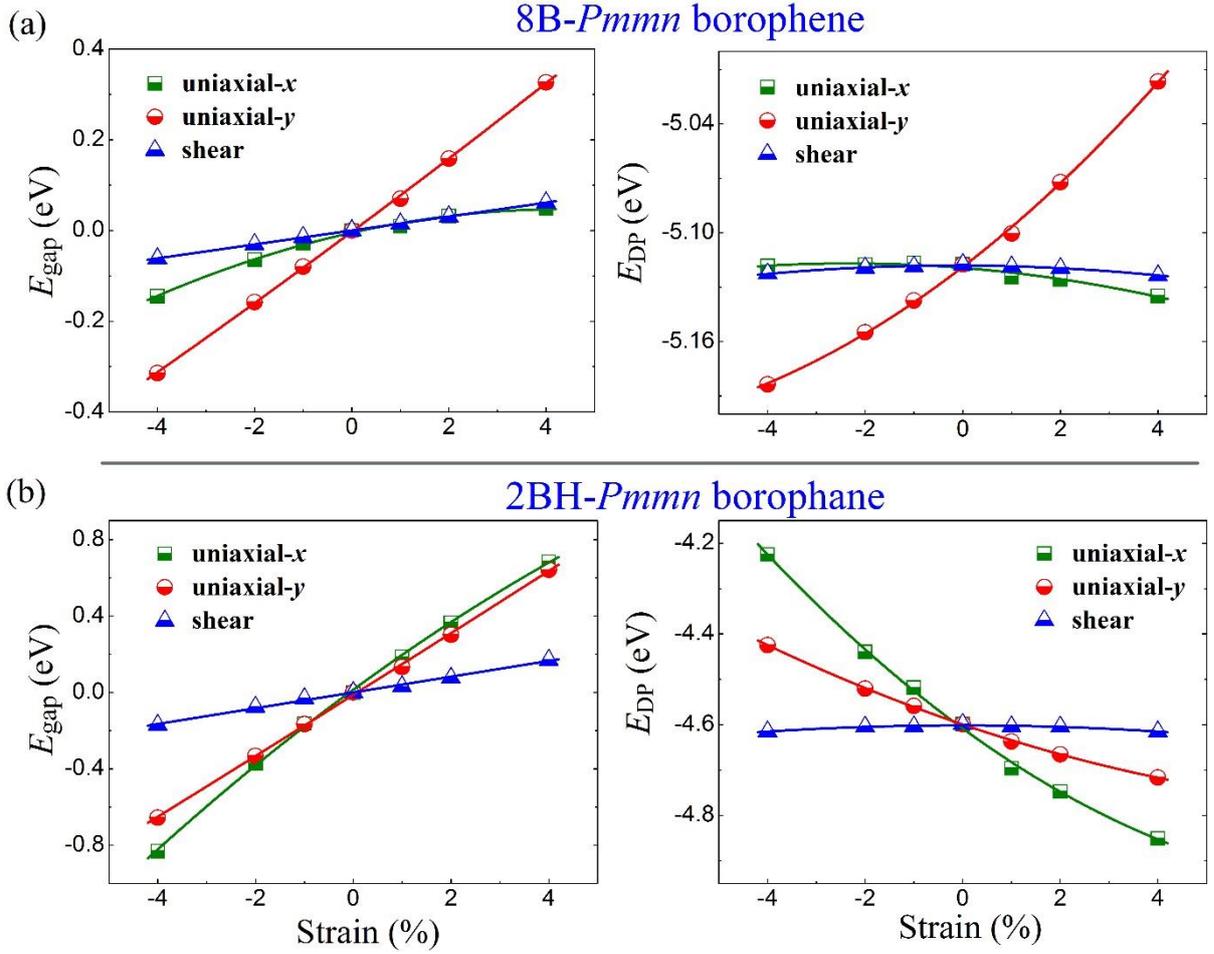

FIG. 4. Determination of the deformation potential constant $E_1$ and the deformation hopping constant $E_\beta$ for (a) 8B-*Pmmn* borophene and (b) 2BH-*Pmmn* borophane. (Left) Induced pseudogap $E_{\text{gap}}$ at the original Dirac point location with respect to three applied strains. (Right) Fermi level (band energy at the Dirac point $E_{\text{DP}}$) as a function of three applied strains. The solid lines are quadratic fits, whose linear coefficients were used to determine $E_1$ and $E_\beta$ based on Eqs. (26) and (27).

The band energy at the Dirac point ($E_{\text{DP}}$, which here is identical to the Fermi level) and the induced pseudogap at the original Dirac point location ($E_{\text{gap}}$) as a function of uniaxial and shear strains are summarized in Fig. 4 for both 8B-*Pmmn* and 2BH-*Pmmn*. For 8B-*Pmmn*, $E_{\text{gap}}$ is perfectly linear with $\varepsilon_{yy}$ and $\gamma$ within the investigated range, but a clear nonlinear effect can be observed for $E_{\text{gap}} \sim \varepsilon_{xx}$. The deformation hopping constant under various strains ($E_{\beta,11}$, $E_{\beta,22}$ and $E_{\beta,66}$) are related to the slopes of $E_{\text{gap}} \sim \varepsilon_{xx}$, $E_{\text{gap}} \sim \varepsilon_{yy}$ and $E_{\text{gap}} \sim \gamma$ via Eq. (27). The obtained data are listed in Table 1 with those of graphene for comparison. Differing from the isotropic $E_{\beta,11} = E_{\beta,22} = E_{\beta,66}$ for graphene, $E_{\beta,66}$ is smaller than $E_{\beta,11}$ and $E_{\beta,22}$ for both 8B-*Pmmn* and 2BH-*Pmmn*. On the other hand, the deformation potential constant $E_1$, and another important parameter $S_1$, are related to the slope of $E_{\text{DP}}$ via Eq. (26). $E_1$ and $S_1$ are zero under shear strain owing to the symmetry. Interestingly, the slope of $E_{\text{DP}} \sim \varepsilon_{yy}$ is positive for 8B-*Pmmn*, making $E_1$ and $S_1$ positive rather than a usual negative, which is the case for most other Dirac materials and semiconductors [43,47]. This may have potential benefits for obtaining a small $|E_1|$ value by hybridizing 8B-*Pmmn* with other materials with negative $E_1$ values, which would result in superhigh carrier mobility.

### D. Carrier mobility

We now consider the carrier mobilities in 8B-*Pmmn* and 2BH-*Pmmn*. In addition to the band structure parameters ($v_{Fx}$, $v_{Fy}$, and $v_t$) and the response parameters of the tilted Dirac cones under strain ($S_{1x}$, $S_{1y}$, $E_{\beta,11}$, $E_{\beta,22}$, and $E_{\beta,66}$), the elastic constants ($C_{11}$, $C_{22}$, and $C_{66}$) are also needed to calculate the mobility. We determined the elastic constants from the total energy variation in first-principles calculations under strain, and their values are listed in Table 1. The results agree well with the literatures [31,33]. The elastic constants of 8B-*Pmmn* are larger than those of 2BH-*Pmmn*. Both of them are anisotropic, where the wrinkling structure along the *x* direction weakens the elasticity, while the strong covalent σ bonds in the *y* direction strengthen the elasticity. With all of the obtained parameters, the mobilities at the neutrality point ($\mu_{\text{NP}}$) and doping condition ($\mu_{doped}$) were calculated using Eqs. (19), (21), and (30)–(33), and the results are summarized in Table 1 and Fig. 5.

**Table 1** Determination of the carrier mobilities.

| System | $\upsilon_{Fx}$ ($10^5$m/s) | $\upsilon_{Fy}$ ($10^5$m/s)[a] | $\upsilon_t$ ($10^5$m/s) | $S_{1x}$ (eV) | $S_{1y}$ (eV) | $E_{\beta,11}$ (eV) | $E_{\beta,22}$ (eV) | $E_{\beta,66}$ (eV) | $C_{11}$ (J/m$^2$)[b] | $C_{22}$ (J/m$^2$)[b] | $C_{66}$ (J/m$^2$)[c] | $\mu_{NP,x/y}$ ($10^5$cm$^2$/V/s)[d] | $\mu_{doped,x/y}$ ($10^5$cm$^2$/V/s)[d] |
|---|---|---|---|---|---|---|---|---|---|---|---|---|---|
| graphene | 8.28 | 8.28 | 0 | −4.3 | −4.3 | 5.2 | 5.2 | 5.2 | 351 | 351 | 144 | 10.4 | 2.40 |
| 8B-*Pmmn* | 5.34 | 7.85 | −3.45 | −0.74 | 0.32 | 1.19 | 3.98 | 0.77 | 253 | 331 | 108 | 14.8, 28.4 | 7.72, 14.9 |
| 2BH-*Pmmn* | 7.70 | 13.48 | −3.86 | −5.16 | −5.96 | 9.39 | 8.05 | 2.06 | 120 | 188 | 37 | 4.48, 13.4 | 0.78, 2.34 |

[a] The Dirac cones in 8B-*Pmmn* and 2BH-*Pmmn* are tilted in the *y* direction, so the final tilted Fermi velocity are $\upsilon_{Fy}+\upsilon_t$ and $\upsilon_{Fy}-\upsilon_t$.

[b] The average values of $C_{11}$ and $C_{22}$ used in the calculations of the mobility.

[c] The elastic constant under shear strain. $C_{66}$ was used according to the Voigt notation instead of $C_{44}$ in previous study[47]. Similarly, $E_{\beta,66}$ was used instead of $E_{\beta,44}$.

[d] The carrier mobilities were calculated with Eqs. (19), (21), and (30)–(33) at room temperature ($T$ = 298 K). $\mu_{doped}$ was determined under the doping condition of $n = 10^{12}$ cm$^{-2}$.

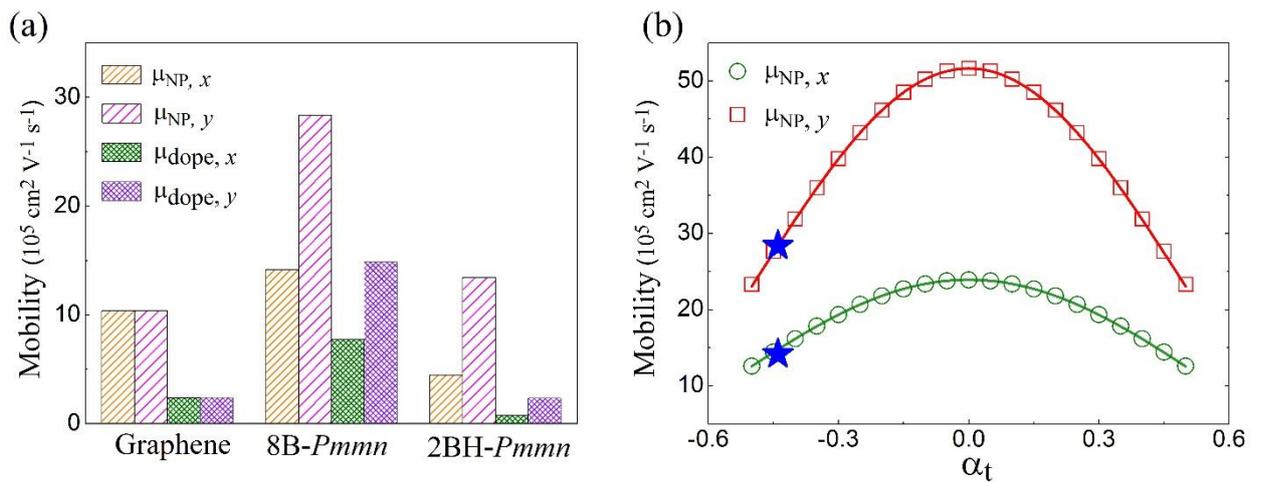

FIG. 5. Carrier mobilities in the considered systems. (a) Comparison of the carrier mobilities in graphene, 8B-*Pmmn* borophene and 2BH-*Pmmn* borophane. $n = 10^{12}$ cm$^{-2}$ was used for the doped systems. (b) The mobility at the neutrality point of 8B-*Pmmn* as a function of the tilting factor $\alpha_t$. The results indicated by symbols were calculated from Eqs. (16) and (18) by numerical integration, while the solid lines were obtained from the analytic formulas in Eqs. (19), (21), (30)–(33). The stars represent the result for 8B-*Pmmn* using the realistic $\alpha_t = -0.44$.

The mobilities in 8B-*Pmmn* and 2BH-*Pmmn* are anisotropic, mainly because of their anisotropic Fermi velocities. In 8B-*Pmmn*, the mobility along the $y$ direction is about two times larger than that in the $x$ direction, and in 2BH-*Pmmn* it is even increased to about three times. The directional-dependent carrier transport makes them more flexible in potential electronic applications. The mobilities in 8B-*Pmmn* at the neutrality point are $14.8 \times 10^5$ and $28.4 \times 10^5$ cm$^2$ V$^{-1}$ s$^{-1}$ for the $x$ and $y$ directions, respectively, both of which are larger than that in graphene ($10.4 \times 10^5$ cm$^2$ V$^{-1}$ s$^{-1}$). At a doping level of $n = 10^{12}$ cm$^{-2}$, the mobility in graphene greatly decreases by about 75% to $2.4 \times 10^5$ cm$^2$ V$^{-1}$ s$^{-1}$, but the mobilities in 8B-*Pmmn* only drop by about 50% to $7.72 \times 10^5$ and $14.9 \times 10^5$ cm$^2$ V$^{-1}$ s$^{-1}$, making its superiority to graphene more prominent [Fig. 5(a)]. For 2BH-*Pmmn*, despite its superhigh Fermi velocity, the mobility is lower than that in 8B-*Pmmn*. This is because scattering by TA phonons dominates the mobility of Dirac cones [47], and thus the larger $E_{\beta,66}$ and smaller $C_{66}$ in 2BH-*Pmmn* are disadvantageous for improving the mobility. This also explains why the mobility of 8B-*Pmmn* is higher than that in graphene.

The analytic formulas for the mobility in Eqs. (19), (21), (30)–(33) were obtained by expanding to the second-order of the tilting factor $\alpha_t$. Alternatively, the mobility can be directly calculated from Eqs. (16) and (18) by numerical integration. In Fig. 5(b), we compare the two methods for $\mu_{NP}$ of 8B-*Pmmn* where $\alpha_t$ is artificially changed. It shows that the discrepancy due to expansion of $\alpha_t$ is very small, [e.g., the error at $\alpha_t = 0.5$ is lower than 0.3%, and the discrepancy in 2BH-*Pmmn* is 2% (data not shown)]. The numerical results are symmetric with respect to $\alpha_t = 0$, which is consistent with the absence of the odd-order terms of $\alpha_t$ in Eqs. (19) and (21). Therefore, the analytic Eqs. (19), (21), (30)–(33) are simple and reliable. In addition, Fig. 5(b) shows that tilting of the Dirac cones has a clear influence on the mobility (i.e., it greatly decreases the mobility). If tilting is removed (i.e., $\alpha_t = 0$), the mobilities in 8B-*Pmmn* at the neutrality point would become $23.9 \times 10^5$ and $51.6 \times 10^5$ cm$^2$ V$^{-1}$ s$^{-1}$ for the *x* and *y* directions, respectively, which are almost twice those at the realistic $\alpha_t = -0.44$. Therefore, it is improper to approximate a tilted Dirac cone with an untilted cone with average Fermi velocity in calculating carrier mobility.

## IV. CONCLUSIONS

In short, we have theoretically studied the intrinsic mobility of materials with tilted Dirac cones. Boltzmann transport theory and deformation potential theory were combined to derive an analytic formula for the mobility under both LA and TA phonons. It indicates that tilting significantly reduces the mobility. The theory was then applied to 8B-*Pmmn* borophene and 2BH-*Pmmn* borophane, both of which have tilted Dirac cones along Γ–X. The predicted undoped mobilities in 8B-*Pmmn* borophene at room temperature are $14.8 \times 10^5$ and $28.4 \times 10^5$ cm$^2$ V$^{-1}$ s$^{-1}$ for the *x* and *y* directions, respectively, both larger than that in graphene. At a doping level of $n = 10^{12}$ cm$^{-2}$, the mobilities are still as high as $7.72 \times 10^5$ and $14.9 \times 10^5$ cm$^2$ V$^{-1}$ s$^{-1}$. For 2BH-*Pmmn* borophane, despite its superhigh Fermi velocity, its mobility is lower than that in

8B-*Pmmn* due to its smaller elastic constant under shear strain. The ultrahigh mobility and anisotropic character make 8B-*Pmmn* borophene promising for potential applications.

## ACKNOWLEDGMENTS

The authors thank Shuqing Zhang for helpful discussions. This work was supported by the National Natural Science Foundation of China (Grant No. 21373015).